# A T-Step Ahead Optimal Target Detection Algorithm for a Multi Sensor Surveillance System

K Madhava Krishna
International Institute of Information Technology
Hyderabad – 500 019, INDIA
Madhav_74_in@yahoo.com, mkrishna@uark.edu

Henry Hexmoor and Shravan Sogani
CSCE Department, University of Arkansas
Fayetteville, AR 72701
hexmoor@uark.edu

*Abstract* – **This paper presents a methodology for optimal target detection in a multi sensor surveillance system. The system consists of mobile sensors that guard a rectangular surveillance zone crisscrossed by moving targets. Targets penetrate the surveillance zone with poisson rates at uniform velocities. Under these conditions we present a motion strategy computation for each sensor such that it maximizes target detection for the next T time-steps. A coordination mechanism among sensors ensures that overlapping and overlooked regions of observation among sensors are minimized. This coordination mechanism is interleaved with the motion strategy computation to reduce detections of the same target by more than one sensor for the same time-step. To avoid an exhaustive search in the joint space of all the sensors the coordination mechanism constrains the search by assigning priorities to the sensors and thereby arbitrating among sensory tasks. A comparison of this methodology with other multi target tracking schemes verifies its efficacy in maximizing detections. A tabulation of these comparisons is reported in the results section of this paper. "Sample" and "time-step" are used equivalently and interchangeably in this paper.**

*Index Terms- multi sensor system, optimal target detection, sensor network, sensor surveillance, multi-agent system*

## I INTRODUCTION

We present a framework for maximizing target detections in a surveillance system consisting of multiple mobile sensors (i.e., robots). The sensors collectively guard a rectangular surveillance zone crisscrossed by moving targets. A-priori information about targets is the knowledge of their poisson rates of entry. Instantaneous or current knowledge about all targets within the sensing range of a sensor is available to the sensor in the form of the target' velocity information and motion direction. While current information is used in deterministically computing the best places to be at in the subsequent time steps, it does not take into consideration the changes in the environment that occur due to consistent flow of target traffic into and out of the surveillance zone. Knowledge of the statistics of target entry and exit on the other hand are used to compute the statistically optimal places to be at in the future from a given location, but this does not consider the current information gleaned through sensor observations. Hence a combination of both current target information and target statistics is used to compute a motion strategy of a sensor that maximizes the number of detections for the next T time steps.

The surveillance zone is discretized into a lattice of cells. Each cell stores information about the expected number of targets to be seen at that cell position purely based on target arrival statistics. This is done offline at all the cell locations across the entire surveillance zone. Then during the real-time phase of the algorithm an estimate of the number of detections based on combination of current target information and target arrival statistics is computed. These values are computed not across the entire lattice but for cells that are at a level T from current positions of all the sensors. Cells that are at a level one from the sensor's current location store detection values for the next T samples, while those that are at level T from the sensor store information only for time step T. By level one we mean the immediate set of eight neighbors in the lattice from the current location of the sensor and level T indicates cells on lattice that are T neighbors away in an eight-connected sense. A depth limited search, with the depth limited by T, maximizing the number of detections performed on the lattice from the sensor's current position yields the motion strategy for the sensor. Computing the best strategy from its individual perspective can result in motions for each sensor with overlapping regions of observation. A priority based coordination mechanism is interleaved with the motion strategy computation to reduce detections of the same target by more than one sensor that further enhances the detection performance for the entire system. To avoid an exhaustive search in the joint space of all the sensors the coordination mechanism constrains the search by assigning priorities to the sensors. Hence while optimal from a single sensor's perspective the algorithm is not optimal across all sensors – a choice that one is reasonable to assume in light of prevailing real time demands. The details of this method are presented in section 3 of this paper. Extensive simulation results confirm the efficacy of this strategy in the form of tabular comparisons presented in section 4.

The proposed approach is particularly suitable for guarding large open areas that are crisscrossed by moving targets when the number of sensors at disposal for monitoring them is limited. Due to limited number of sensors as well as to glean characteristics of targets over several observations the sensors are allowed to be move. Such surveillance systems find utilities in many security, surveillance and reconnaissance applications.

The rest of the paper is organized as follows. Section 2 places the current work in the context of similar works found in the literature. Section 3 presents the formulation of the methodology and section 4 attests to the efficacy of the methodology in our simulation work. Section 5 concludes and provides further scope for this work.

## II. BACKGROUND REVIEW

Many security, surveillance, and reconnaissance applications require distributed autonomous observation of movements of targets navigating in a bounded area of interest. Multi-sensor surveillance finds applications such as in border patrol, guarding of secured areas, search and rescue and warehouse surveillance [1, 2]. It involves detection of multiple intrusions and/or tracking through coordination between the sensors. Detection and target tracking has been researched from multiple viewpoints. Some efforts have focused on the problem of identifying targets from a given set of data through particle filters [3], and probabilistic methods [4]. The problem of data association or assigning sensor measurements to the corresponding targets were tackled by Joint Probabilistic Data Association Filters by the same researchers such as in [3]. Kluge and others [5] use dynamic timestamps for tracking multiple targets. Krishna and Kalra [6] presented clustering based approaches for target detection and further extended it to tracking and avoidance. The focus of these approaches has been on building reliable estimators for predicting target trajectories that is different from the objective of this effort to maximize target detections.

In the context of distributed task allocation and sensor coordination Parker proposed a scheme for delegating and withdrawing robots to and from targets through the ALLIANCE architecture [7]. The protocol for allocation was one based on "impatience" of the robot towards a target while the withdrawal was based on "acquiescence". Jung and Sukhatme [8] present a strategy for tracking multiple intruders through a distributed mobile sensor network and a strategy for maximizing sensor coverage[8, 9]. Lesser's group have made significant advances to the area of distributed sensor networks [10] and sensor management [11]. In [12] Parker presents a scheme called A-CMOMMT where the goal is to maximize the number of targets observed over a time interval of length $T$ based on the same philosophy of behavior-based control as in [7]. The authors of this paper present their scheme for resource allocation and coordination in a distributed sensor system through a set of fuzzy rules in [13] and further compare various resource allocation strategies in terms of their detection performance in [14]. The author of [15] has looked at the problem of static placement of sensors in known polygonal environments and [16] describes a distributed sensor approach to target tracking using fixed sensor locations. The current approach is disparate from those of [15,16] in that in the current scheme the sensors are mobile.

Among the approaches that we have encountered the closest to this effort are [12] and [8]. In [12] a behavior-based approach, A-CMOMMT, is compared with three other heuristic approaches where the sensor's motion strategy is arbitrary or random in the first, stationary (the sensor does not move) in the second and based on local force control in the third. However the approaches including A-CMOMMT are heuristic in that while effective they do not guarantee that the motion strategy generated is actually the best possible strategy for those T time steps based on the gathered data. In [8] a motion strategy for tracking multiple targets based on density estimates is presented. The robot attempts to maximize target detections by maintaining itself at a particular distance from the center of gravity of currently observed targets. It is nonetheless not clear if this could again be the best strategy.

In contrast, our approach presents a scheme that moves the sensor to provably the optimal regions in the surveillance region in order to maximize detections over the next T samples. In other words, through a depth limited search the sensor comes up with the best possible motion strategy that results in maximizing detections over next T time steps among the various possible strategies from the current sensor location. This is combined with the coordination mechanism to reduce overlapping areas of observation positions this method as novel and different from other approaches the authors have come across in literature.

## III. THE METHODOLOGY

### A. Description of Surveillance Zone, Sensors and Targets:

We consider the surveillance system depicted in figure 1. The sides of the outer rectangle or the biggest rectangle in figure 1 form the boundary of the surveillance zone – the area enclosed by it is the area of interest where sensors attempt to optimize their rates of detection. The shaded circles are effective sensor ranges in their starting positions. The field of vision (FOV) of a sensor is 360 degrees. The squares with thick boundaries in figure 1 with the sensors at their center are the inscribed squares of the circular FOV of a sensor. In other words, the diameter of the FOV of a sensor is the diagonal of one such inscribed square. Purely from the point of view of facilitating easier computations the sensor considers only those targets that lie within its inscribed square as targets within its FOV. It needs to be emphasized that this simplification does not have any bearing on the overall philosophy of this approach. In the results section the efficacy of this method is verified by uniformly applying this same condition across all other approaches that are compared and the extension to a case of circular FOV is merely one of more involved but computable computations. The entire surveillance region is discretized into a lattice of cells. The cells are represented as the small squares inside the FOV of the leftmost and topmost sensor. The dashed lines along the length and breadth indicate that the cells proceed to fill the entire surveillance zone. At each of the cell locations various aspects of target statistics are computed that are described later. The crosses outside the surveillance zone are the source points from where targets emanate as per Poisson statistics. Targets percolate from each of those sources into that horizontal or vertical half-plane that contains the surveillance zone. Therefore, all targets coming from a particular source will be contained with an angular span of $\pi$ radians. Furthermore, the following assumptions are made for sensors and targets

- A sensor can detect all targets within its FOV or occlusion relations are not considered.
- The takeoff angle of a target from its source point is uniformly distributed in $[0, \pi]$
- All targets move with the same uniform velocity within the surveillance zone along linear trajectories, which can be ascertained by the sensor.

The last assumption allows that the statistical values of various parameters computed at every cell in the lattice to take a unique value rather than a probability distribution. In case of a distribution the expected values of the parameters need to be made use of.

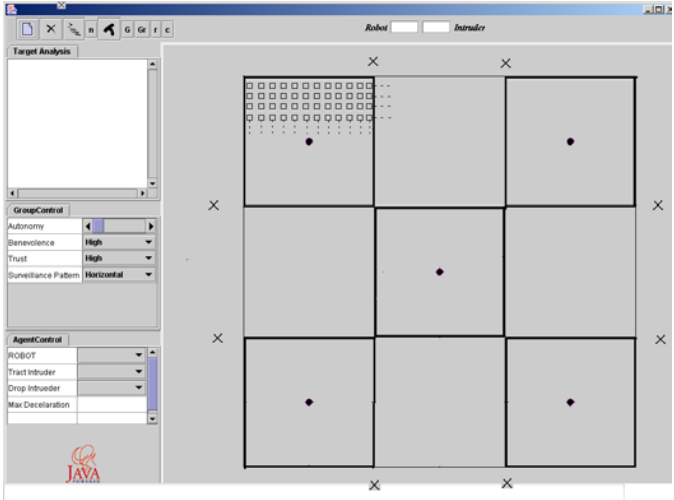

Figure 1: The surveillance zone is represented by the outer rectangle. The sensors by shaded circles, the crosses outside are the target sources and the small squares within the FOV of top and left sensor are the discretized cells. The dashed lines along the length and breadth indicate that the cells proceed to fill the entire surveillance zone.

B  Problem Statement and Approach

The problem attacked in this paper is stated as follows. Given:
- $N_S$: The set of all sensors in the system, $N_S = \{s0, s1,..., sn_s\}$, where $si$ denotes the sensor with label $i$, ordered in a sequence. Hence $n_s$ the label of the last enumerated sensor is also the number of sensors in the system, which is a constant.
- $N_{I,t}$: The set of all targets in the system at time $t$, where, $N_{I,t} = \{i0, i1,..., in_{I,t}\}$, and $n_{I,t}$ is the number of targets in the system at time step $t$ that varies at every time step and hence dynamic.
- $g_{m,t}$: A binary variable that takes the value 1 if a target $im$ is observed by any one of the $n_s$ sensors at $t$.

Objective: To develop an algorithm such that the following cost function $J = \sum_{t=1}^{T} \sum_{m=1}^{n_{I,t}} g_{m,t}$ is maximized. In other words, the number of detections of the targets present in the system at every time step is maximized over T time steps.

*Approach*: The methodology is delineated as a sequence of steps below and some of the steps are elaborated therein and some are explained subsequently. Due to brevity of space, some of the computations mentioned in these steps that peripheral to the core algorithm are not detailed.

*1)* At each cell, $Pi$, in the lattice the following are computed and stored.

*1a)*. $\hat{\lambda}_{Pi}$: The expected rate of target entry into the FOV of a sensor centered at $Pi$. If there are $Q$ target sources each emanating targets at rate $\lambda$ then $\hat{\lambda}_{Pi} = \lambda \sum_{j=1}^{Q} \frac{\theta_j}{\pi}$, where the ratio $\frac{\theta_j}{\pi}$ denotes the expected fraction of the total number of targets that would enter the FOV of the sensor at $Pi$ from the $j^{th}$ of the $Q$ target sources. In figure 1 $Q=8$. We also denote by $\lambda_{Pi,j} = \lambda \frac{\theta_j}{\pi}$ the rate at which the sensor at $Pi$ sees targets from source $j$. In other words, among the targets taking off at any angle within $[0, \pi]$, from source $j$ only those that takeoff within the angular span $[\delta_j, \delta_j + \theta_j]$ would enter the FOV of the sensor, where $\delta_j$ is the smallest angle formed by the segment connecting a vertex of the FOV square and source $j$ and $\delta_j + \theta_j$ is the largest angle.

*1b)* $\hat{Te}_{Pi}$: The expected escape time for a target through the field of vision of the sensor centered at Pi. The escape time is essentially the time for which a target would be in the FOV before it escapes from a sensor.

*1c)* $\hat{d}_{Pi} = \hat{\lambda}_{Pi} \hat{Te}_{Pi}$: The number of targets expected to be detected by a sensor at $Pi$ for any time step.

*1d)* $\eta_{Pi,j}$: The normalization constant at cell $Pi$ due to target source $j$. This constant renders the summation of the probability distribution function (pdf) of the escape time of a target through the FOV of a sensor at $Pi$ due to source $j$ to go to unity. In other words if $te_{Pi,j}$ is the random variable that measures the escape time of a target for a sensor at cell $Pi$ due to source $j$ and $f(te_{Pi,j})$ be its pdf then $\eta_{Pi,j}$ sees to that $\int_{t_a}^{t_b} f(te_{Pi,j}) = 1$, where the lower and upped bounds of the integral are the minimum and maximum possible escape times. It is to be noted that in general $f(te_{Pi,j})$ is not the same in $[t_a, t_b]$ and needs to be broken into sub functions for respective time intervals for which it has the same closed form representation. The details of how these sub probability distribution functions are computed is not given here for conciseness.

2) For every sensor $sj$ in the system located at a particular cell location $Pj$ the path it takes for the next T time steps is computed such that the number of target detections is maximized based on steps 3 and 4.

3) For $sj$ currently at $Pj$ compute for all cells, $Pk$, that are at level $n$, $n = \{0,1,2,...,T\}$ from $Pj$, the expected number of detections, $\hat{n}_d^{k,t}$, where the first index, $k$, in the superscript refers to the cell $Pk$ and the second index $t$ denotes the time step for which it was computed. The computation is based on what $sj$ currently observes at $Pj$ and the target arrival statistics, elements of which are computed in step 1. This computation is carried out at the current cell $Pj$ and those cells one time-step away for the next T time steps. This is done because the sensor can choose to remain at the current cell itself for the next T time steps or migrate to a neighboring cell and stay there for next T-1 steps or come back to the starting cell for any of the future T steps depending on whichever of those various possible strategies yields maximum detections. For cells at depth two this computation is carried T-1 times, corresponding to time steps starting from 2 and ending at T with T inclusive. For a cell at depth T the computation is done once for time step T alone. This results in a tree where each cell at time step $t$ is connected to its eight nearest cells as well as itself at $t+1$. Since the cell that connects to itself does for a different time step the two cells though at the same location are different due to difference in time as their information content varies with every time step. Figure 2 shows an instance of the tree, where a cell location labeled as $(i, j, t)$ connects to its eight neighbors as well as itself at $t+1$, where $(i, j)$ represents the index of the cell in the two dimensional lattice and $t$ its index in time. Thus every node in the tree except the leaf node is connected to nine nodes and the depth of the tree is T.

4) On this nine connected space-time tree perform a depth-limited search to maximize the detections with $\hat{n}_d^{k,t}$ as the objective value of a link that connects a current cell node $Pj$ at $t$-1 to another node $Pk$ for the next time step $t$.

5) The paths of the sensors are ordered based on the number of detections. The sensor path that has the highest number of detections has the highest priority. Ties are broken randomly.

6) For overlapping FOV between sensors at certain locations in the paths the number of detections is reduced at the corresponding cell for the corresponding time-step for the sensor with lower priority. The reduction is computed as $\hat{n}_d^{k,t} = \hat{n}_d^{k,t} - f\,\hat{n}_d^{k,t}$. The fraction $f$ is proportional to the area of overlaps given by the area of overlapping FOV divided by the area of the FOV.

7) The optimal path for the sensor with the lower priority is recomputed based on the updated $\hat{n}_d^{k,t}$ values through depth limited search. For a sensor with priority $p$, we check for overlaps with all those sensor trajectories with priorities 1, 2, ..., p-1.

8) The steps two to seven are repeated until the end of simulation.

Step 1 of the algorithm is performed only once at the start of the simulation across all the cells and is essentially an offline step unless the target statistics changes dynamically.

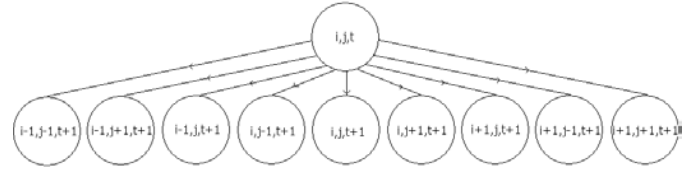

Figure 2: An instance of the nine connected spacetime tree, where every node except the leaves connects to its eight neighbors as well as itself at the next time step.

B Computing $\hat{n}_d^{k,t}$

Computation of $\hat{n}_d^{k,t}$ varies if it is being carried out for the same cell location for a future time or for a different cell location. For the case of computing at the same cell the procedure is as follows:

Given that the sensor at cell $Pi$ currently or at $t=0$ observes $n_{Pi}$ number of targets then the number of targets it is likely to see $T$ time-steps into the future is given by:

$$\hat{n}_d^{i,T} = n_{Pi} - k + \hat{\lambda}_{Pi} T - \sum_{j=1}^{Q} \hat{\lambda}_{Pi,j} \sum_{t=1}^{T} P(te_{Pi,j} < T - t)$$

(1)

Here $\hat{\lambda}_{Pi}$, $te_{Pi,j}$, $\lambda_{Pi,j}$ have the same connotations as discussed in step 1 and $Q$ is the number of target sources as before. The first term on the right hand side or RHS of equation (1) is the deterministic part that is computed purely based on what the sensor senses now. It merely states that out of $n_{Pi}$ particle seen currently $k$ of them would disappear by $T$ time steps. The second term onwards on the RHS denote the statistical counterpart. It says that $\hat{\lambda}_{Pi} T$ are likely to enter the FOV in $T$ time steps and of which the number in the third term are likely to leave the FOV by $T$. The third term containing the summation is explained as follows. $P(te_{Pi,j} < T - t)$ denotes the probability that a target from source $j$ that entered the FOV at $t$ would have escaped the FOV by $T$. Hence $\hat{\lambda}_{Pi,j} P(te_{Pi,j} < T - t)$ is the estimate of the number of particles that entered at $t$ and escaped by $T$, since $\hat{\lambda}_{Pi,j}$ is the estimate of number of particles entering for every $t$ from $j$ into the FOV of the sensor at $Pi$. The summation over $t$ signifies that estimate of the number of particles that would have left by $T$ needs to be done in correspondence with the time at which they entered between now and $T$. The summation over $Q$ signifies that the pdf of $te_{Pi,j}$ is not the same for every source $j$ at that cell location. The $P(te_{Pi,j} < T-t)$ is given below by its pdf with notations from step 1.

$$P(te_{Pi,j} < T-t) = \begin{cases} 0 & ; T-t < t_a \\ \frac{1}{\eta_{Pi,j}} \int_{t_a}^{T} f(te_{Pi,j}); t_a \leq T-t \leq t_b \\ 1 & ; T-t > t_b \end{cases}$$

The number of candidates likely to be seen for a different cell $Pk$ at $T$ given that $n_{Pi}$ number of targets then the number of targets are currently seen at $Pi$ is given by:

$$\hat{n}_d^{k,T} = n_{Pi} - k + \kappa \left( \hat{\lambda}_{Pi} T - \sum_{j=1}^{Q} \hat{\lambda}_{Pi,j} \sum_{t=1}^{T} P(te_{Pi,j} < T-t) \right)$$
$$+ (1-\kappa) \hat{d}_{Pk} \qquad (2).$$

Equation (2) has similar connotations as (1) except for the appearance of $\kappa$ that represents the fraction of area common to FOV erected at $Pi$ and $Pk$ and $\hat{d}_{Pk}$ is as discussed in step 1. For that fraction of the FOV at $Pk$ that is not visible from $Pi$, $\hat{d}_{Pk}$, which is the expected number of detections at cell $Pk$ at any instant is made use of. And since $(1-\kappa)$ is the fraction of FOV at $Pk$ that is not visible $\hat{d}_{Pk}$ is multiplied by that fraction. This makes use of the assumption that the distribution of targets is uniform across an area. In our plots of $\hat{d}_{Pk}$ done for the surveillance zone variations between cells are steep only towards the edges of the zone. A more rigorous computation involves evaluating the pdf for the overlapped and non-overlapped areas separately. This is being avoided since (2) is computed during the online phase of the algorithm. This assumption is also invoked in step 6 while reducing the number of detections due to overlapping FOV.

*C The Coordination Phase:*

The coordination phase prevents paths of sensors to come close to one another to avoid overlapping FOV. This is done by imposing penalties and reducing values of $\hat{n}_d^{k,t}$ and recomputing paths for sensors with lower priorities as described in steps 5, 6 and 7. The path of the sensor with highest detections is fixed. The path with the second highest recomputed if there are overlapping areas of observation with the first. The recomputed path constitutes a motion strategy that is optimal in terms of detections under the constraint that the path of highest sensor is fixed. Similarly the path of the sensor with the least priority sensor when recomputed is optimal under the constraint that the paths of the sensors with higher priority are fixed. Hence at the coordination phase the optimality of the algorithm is not complete. A fully complete algorithm would involve a search in the joint space of all sensors that is combinatorially hard. The depth limited search over T=3 time steps involves a search over $9^3 = 729$ paths for reaching an optimal motion strategy. Hence a search in the joint space for $n$ sensors involves a search over $729^n$ paths, that is computationally infeasible for online applications. The coordination phase resembles the decoupled priority based approaches to multi robot path planning [17,18].

IV SIMULATION RESULTS

In this section we report results obtained through simulations on our environment developed using Borland's JBuilder IDE. The value of the number of time steps, T, used for these simulations is 3, carried out on a P4 workstation with clock speed of 1.8 GHz. Figure 3 shows a snapshot of the environment with 10 sensors. The current position of the sensor is shown through the bigger shaded circle, while the targets through smaller ones. The traces of sensor movements are also shown. Target traces are not shown to avoid cluttering.

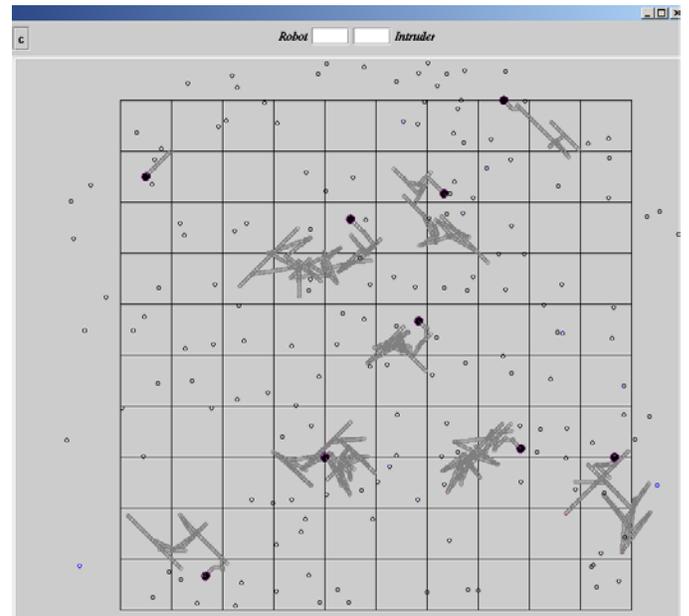

Figure 3: A simulation system with ten sensors and several targets. The robots are shown through larger circles and targets through smaller ones. The traces of sensors are also shown.

Table 1 compares the performance of the system in the presence and absence of coordination. The experiments done with sensor coordination are labeled as 1a, 2a, …, while those without coordination are labeled as 1b, 2b,… . Each experiment lasted for 150 time steps in total. The first column of the table denotes the index of the experiment; the second denotes the number of sensors in the system, the third the average detections or the number of targets detected by at-least one sensor per sample. The fourth signifies the average fraction of targets detected per sample, which is the number of targets detected per sample by one or more sensors divided by the number of targets in the surveillance zone per sample. Columns 5, 6 and 7 represent the number of targets detected exactly by one, two and three sensors per sample. The last column denotes the target velocity. Detections by more than three sensors are not tabulated for they generally assume insignificant values. Of particular interest in comparing the two schemes are the average fraction of target detections as well as the number of targets detected by exactly one, two and three sensors. It is expected that the average fraction of detections as well as the number of targets detected by exactly one sensor to be more for the method with coordination incorporated. Simultaneously the number of targets detected by two or more sensors is anticipated to be higher for the method that does not use the coordination phase. As the overlapping areas of FOV are higher in the absence of coordination the number of targets detected by two or more sensors is also higher. This in turn allows more targets to go undetected and the average fraction of detections to be lesser. The abbreviations used in the column headings are explained in the table caption.

| EN | NS | AD | ZD | AF | D1S | D2S | D3S | TV |
|----|----|----|----|----|-----|-----|-----|----|
| 1a | 10 | 32 | 22 | 0.6 | 29 | 2 | 0 | 10 |
| 1b | 10 | 27.1 | 24.9 | 0.52 | 19.5 | 6.8 | 0.8 | 10 |
| 2a | 6 | 22.2 | 25.7 | 0.46 | 22 | 0.2 | 0 | 10 |
| 2b | 6 | 21.5 | 26.8 | 0.44 | 18.4 | 3.2 | 0 | 10 |
| 3a | 10 | 20.4 | 21.1 | 0.49 | 20 | 0.3 | 0 | 15 |
| 3b | 10 | 17.3 | 24.1 | 0.41 | 13.8 | 3.5 | 0 | 15 |
| 4a | 6 | 27.2 | 20.2 | 0.57 | 26.5 | 0.7 | 0 | 10 |
| 4b | 6 | 26.9 | 20.4 | 0.56 | 25.7 | 1.3 | 0 | 10 |
| 5a | 10 | 20 | 22.5 | 0.47 | 19.6 | 0.4 | 0 | 15 |
| 5b | 10 | 18.7 | 23.8 | 0.44 | 16.4 | 1.5 | 0.9 | 15 |

Table 1: Comparison between motion strategies with and without sensor coordination. Experiments with labels ending in 'a' or the first line in every row have coordination incorporated while those ending in 'b' or the second line in each row are uncoordinated. Abreviations are as follows: EN = experiment number, NS = number of sensors, AD = Average Detections per sample, ZD = Zero Detections per sample, AF = Average Fraction per sample, D1S = Detections by exactly 1 sensor, D2S = Detections by exactly 2 sensors, D3S = Detections by exactly 3 sensors, TV = Targets' Velocity

As seen from table 1 the performance of the coordinated scheme with decoupled optimization is better for all experiments with differing NS and TV values than the method sans coordination. For example the number of targets detected by exactly one sensor per sample, the D1S column, is significantly higher in a number of experiments for the coordinated vis-à-vis the uncoordinated scheme. It is higher by 10, 4 and 6 detections per sample in experiments 1, 2 and 3. Also the number of targets detected by two sensors is significantly lesser across all experiments for the coordinated method. In experiments 1 and 5 the number of targets detected by three sensors is almost one per sample for the method lacking coordination. Correspondingly average fraction of targets detected per sample for the coordinated method is higher in all the experiments. The results of the table tally with the expectations and the underlying reasons for these expectations mentioned earlier.

In table 2 we compare the current strategy with four of our previous methods of target detection and pursuit [14]. The first column signifies the experiments. Indexes 1a, 2a, etcetera correspond to the current method. Those with indices 'b', 'c', 'd' and 'e' correspond to coordinated-distracted, coordinated-dedicated, local-distracted and local-dedicated methods of resource allocation of sensors to targets for target detection [14]. Table 2 shows that across all experiments the current scheme surpasses all others in terms of average fraction of targets detected as well as other criteria such as D1S, D2S, AD and AF. It is to be noted that a difference of 0.06 in average fraction per sample between two methods corresponds to a difference of 4 or 5 target detections per sample (there are 70 to 80 targets in the zone per time-step on an average), which in turn corresponds to a difference of 600 detections in 150 time-steps of simulation.

| EN | NS | AD | ZD | AF | D1S | D2S | D3S | TV |
|----|----|----|----|----|-----|-----|-----|----|
| 1a | 6 | 22.3 | 25.7 | 0.46 | 22 | 0.2 | 0 | 10 |
| 1b | 6 | 20.3 | 27.7 | 0.42 | 17.9 | 2.3 | 0.1 | 10 |
| 1c | 6 | 18.3 | 30.5 | 0.37 | 13.1 | 4.4 | 0.8 | 10 |
| 1d | 6 | 21 | 27 | 0.43 | 18.7 | 2.3 | 0 | 10 |
| 1e | 6 | 18.3 | 30.5 | 0.37 | 13.1 | 4.6 | 0.5 | 10 |
| 2a | 10 | 21.5 | 20.2 | 0.46 | 21.0 | 0.5 | 0.0 | 15 |
| 2b | 10 | 19.0 | 26.1 | 0.40 | 17.5 | 1.6 | 0.0 | 15 |
| 2c | 10 | 19.7 | 25.8 | 0.42 | 18.0 | 1.6 | 0.02 | 15 |
| 2d | 10 | 19.2 | 25.4 | 0.41 | 17.0 | 2.2 | 0.02 | 15 |
| 2e | 10 | 20.0 | 25.8 | 0.42 | 18.8 | 1.2 | 0.0 | 15 |
| 3a | 6 | 27.2 | 20.2 | 0.57 | 26.5 | 0.7 | 0 | 10 |
| 3b | 6 | 21.2 | 26.1 | 0.44 | 20.2 | 1.0 | 0 | 10 |
| 3c | 6 | 21.5 | 25.8 | 0.45 | 19.8 | 1.7 | 0 | 10 |
| 3d | 6 | 21.9 | 25.4 | 0.46 | 20.9 | 1.0 | 0 | 10 |
| 3e | 6 | 21.5 | 25.7 | 0.45 | 19.8 | 1.7 | 0 | 10 |
| 4a | 10 | 40.4 | 66.0 | 0.40 | 39.8 | 0.7 | 0.0 | 10 |
| 4b | 10 | 30.0 | 68.9 | 0.32 | 26.3 | 3.5 | 0.2 | 10 |
| 4c | 10 | 30.9 | 74.4 | 0.31 | 29.0 | 1.9 | 0.0 | 10 |
| 4d | 10 | 31.6 | 73.7 | 0.31 | 27.4 | 3.5 | 0.8 | 10 |
| 4e | 10 | 31.6 | 73.8 | 0.30 | 29.8 | 1.8 | 0.0 | 10 |
| 5a | 6 | 30.0 | 57.4 | 0.35 | 29.3 | 0.7 | 0.0 | 15 |
| 5b | 6 | 21.4 | 64.9 | 0.24 | 17.7 | 3.4 | 0.2 | 15 |
| 5c | 6 | 25.9 | 60.3 | 0.30 | 22.5 | 3.4 | 0.04 | 15 |
| 5d | 6 | 24.8 | 61.4 | 0.28 | 21.0 | 3.7 | 0.04 | 15 |
| 5e | 6 | 25.5 | 60.7 | 0.29 | 22.4 | 3.2 | 0.0 | 15 |

Table 2: Comparison of current method (label a) with four previous methods called coordinated-distracted (b), coordinated-dedicated (c), local-distracted (d) and local-dedicated (e). The abreviations are same as in table 1.

The comparisons reported in the table were done by introducing targets according to Poisson rates for one of the methods for an experiment that we call as the base method. For the remaining methods of that experiment the targets were introduced in the same fashion as the base method. The value of $\lambda$ was 0.3 and $Q$=8 in these simulations.

V CONCLUSIONS AND SCOPE

A method for motion strategy computation of a sensor that maximizes the number of target detections for the next T-

time steps is presented. The method makes use of a-priori known statistics of target arrivals along with detections reported for the current sample to estimate the number of detections on a lattice of cells. To reduce overlaps a coordination mechanism is specified that performs a constrained search, where the constraints are in the form of priorities assigned to sensors. The absence of an exhaustive search in the joint space due to the constraints renders the algorithm sub-optimal from a multi sensor perspective although optimal with respect to a single sensor in presence of those constraints. The tabulations presented in the results section vindicate the performance of the current approach in comparison to previous approaches for target detection and pursuit. The performance enhancement due to the coordination phase that reduces overlaps in FOV between sensors is also tabulated in section IV.

As with any optimization algorithm the tradeoff is between the completeness of the optimization method vis-à-vis real-time requirements. In this aspect considering real time requirements a greedy decoupled optimization is resorted to at the multi sensor level. The performance is also dependent on the resolution of the discretization or the distance between the cells in the lattice.

Future scope of this effort lies in identifying possible geometric properties of the surveillance area and nature of sensor paths to reduce the search space in the joint space of sensors that would provide for a real-time multi-sensor optimally detecting algorithm. Other areas of interest involve modifying the framework to provide for probability distributions for statistical parameters computed at the cells that allow targets to take multiple velocities. We also intend to compare the current method with other known methods [12] and extend the current method to include obstacles of simple geometry within surveillance area.